# Attribute oriented induction with star schema

Spits Warnars H.L.H
Department of Computing and Mathematics ,Manchester Metropolitan University
John Dalton Building, Chester Street, Manchester M1 5GD, United Kingdom
s.warnars@mmu.ac.uk, Telp. +44 (0)161 247 1779, fax: +44 (0)161 247 6831

Abstract :
This paper will propose a novel star schema attribute induction as a new attribute induction paradigm and as improving from current attribute oriented induction. A novel star schema attribute induction will be examined with current attribute oriented induction based on characteristic rule and using non rule based concept hierarchy by implementing both of approaches. In novel star schema attribute induction some improvements have been implemented like elimination threshold number as maximum tuples control for generalization result, there is no ANY as the most general concept, replacement the role concept hierarchy with concept tree, simplification for the generalization strategy steps and elimination attribute oriented induction algorithm. Novel star schema attribute induction is more powerful than the current attribute oriented induction since can produce small number final generalization tuples and there is no ANY in the results.

1. Introduction

DBLearn is a prototype data mining system which developed in Simon Fraser University integrates machine learning methodologies with database technologies and efficiently and effectively extracts characteristic and discriminant rules from relational databases (Han et al. 1994; Han, Fu and Tang 1995a). Since 1993 DBLearn have led to a new generation of the system call DBMiner with the following features:
   (1) Incorporating several data mining techniques like attribute oriented induction, statistical analysis, progressive deepening for mining multiple-level rules and meta-rule guided knowledge mining (Han et al. 1996) data cube and OLAP technology (Han et al 1997).
   (2) Mining new kinds of rules from large databases include multiple level association rules, classification rules, cluster description rules and prediction.
   (3) Automatic generation of numeric hierarchies and refinement of concept hierarchies.
   (4) High level SQL-like and graphical data mining interfaces.
   (5) Client server architecture and performance improvements for larger application.
   (6) SQL-like data mining query language DMQL and Graphical user interfaces have been enhanced for interactive knowledge mining.
   (7) Perform roll-up and drill-down at multiple concept levels with multiple dimensional data cubes.

DBMiner had been developed by integrating database, OLAP and data mining technologies (Han et al. 1997) which previously called DBLearn have database architecture as shown in figure 1. Concept hierarchy is stored as a relation in the database provides essential background knowledge for data generalization and multiple level data mining. Concept hierarchy can be specified based on the relationship among database attributes or by set groupings and be stored in the form of relations in the same database (Han et al. 1996). Concept hierarchy can be adjusted dynamically based on the distribution of the set of data relevant to the data mining task and hierarchies for numerical attributes can be constructed automatically based on data distribution analysis (Han et al. 1996).

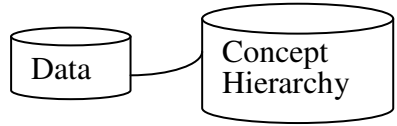

Figure 1. DBMiner database architecture

2. Problem Definition

Attribute oriented induction has a weakness where it only provides a snapshot of the generalized knowledge and not a global picture and global picture in attribute oriented induction can be revealed by trying different thresholds repeatedly. As result by setting different thresholds will obtain different sets of generalized tuples and using different thresholds repeatedly is a time consuming and tedious work (Wu et al 2009). Based on this weakness a novel approach for attribute induction has been proposed where thresholds number as a control for maximum number of tuples of the target class in the final generalized relation will no longer be needed and will be replaced with group by operator in sql select statement.

In the proposed star schema attribute induction the role of concept hierarchy will be substituted by concept trees where concept tree as concept hierarchy simplification (Cheung et al. 2000) and represent taxonomy of concepts of the values in an attribute domain (Han et al, 1992; Han et al, 1995b). Figure 2 show star schema database architecture for attribute induction, where amount of concept tree table will depend on how many concept tree which is degraded from concept hierarchy. In other word the amount of concept tree table will represent the amount of concept tree.

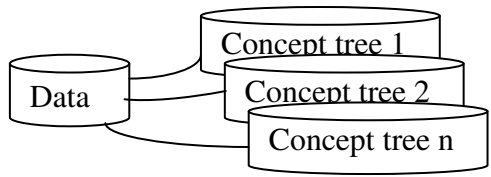

Figure 2. Star schema attribute induction database architecture

In current attribute oriented induction, query is processed with SQL-like data mining query language DMQL in the beginning process for collecting the relevant set of data by processing a transformed relational query, generalizes the data by attribute oriented induction and then presents the outputs in different forms (Han et al. 1996). In the star schema attribute induction query process is not just only prepare for collecting the relevant set of data in the beginning but query process will be enhanced to generalizes the data by attribute oriented induction based on the star schema database architecture.

For the next section the novel approach star schema attribute induction will be examined by the implementation the current attribute oriented induction and star schema attribute induction and the examination will just only focus on characteristic rule and using non rule based concept hierarchy. The implementation will be done with Java programming language for making the application and Mysql database for the repository. For convenience, easy to learn and connectivity with current research both of approaches will be implemented with the same data and concept hierarchy based on data example in Cai (1989) and Han et al. (1992). The application will be run on the computer with specification Mobile Intel Pentium 4, 2.20 GHz, 644 MHz and 512 MB of RAM.

3. Data and Concept hierarchy

In this section will declare the data example and concept hierarchy refer to data example from Cai (1989) and Han et al (1992) which will be used for examination characteristic rule between current attribute induction and star schema attribute oriented induction. Table 1 is an example of graduate student data which was adopted from Cai (1989) and Han et al. (1992) and figure 3 is a non rule based concept hierarchy and concept trees all at once which also was adopted from Cai (1989) and Han et al. (1992).

| Name | Category | Major | Birthplace | GPA |
|---|---|---|---|---|
| Anton | M.A. | History | Vancouver | 3.5 |
| Anil | M.S. | Physics | Ottawa | 3.9 |
| Ayin | Ph.D. | Math | Bombay | 3.3 |
| Abdi | Ph.D. | Biology | Shanghai | 3.4 |
| Agung | Ph.D. | Computing | Victoria | 3.8 |
| Ahing | M.S. | Statistics | Nanjing | 3.2 |

Table 1. Graduate student data

International Journal of Database Management Systems ( IJDMS ) , Vol.2, No.2, May 2010

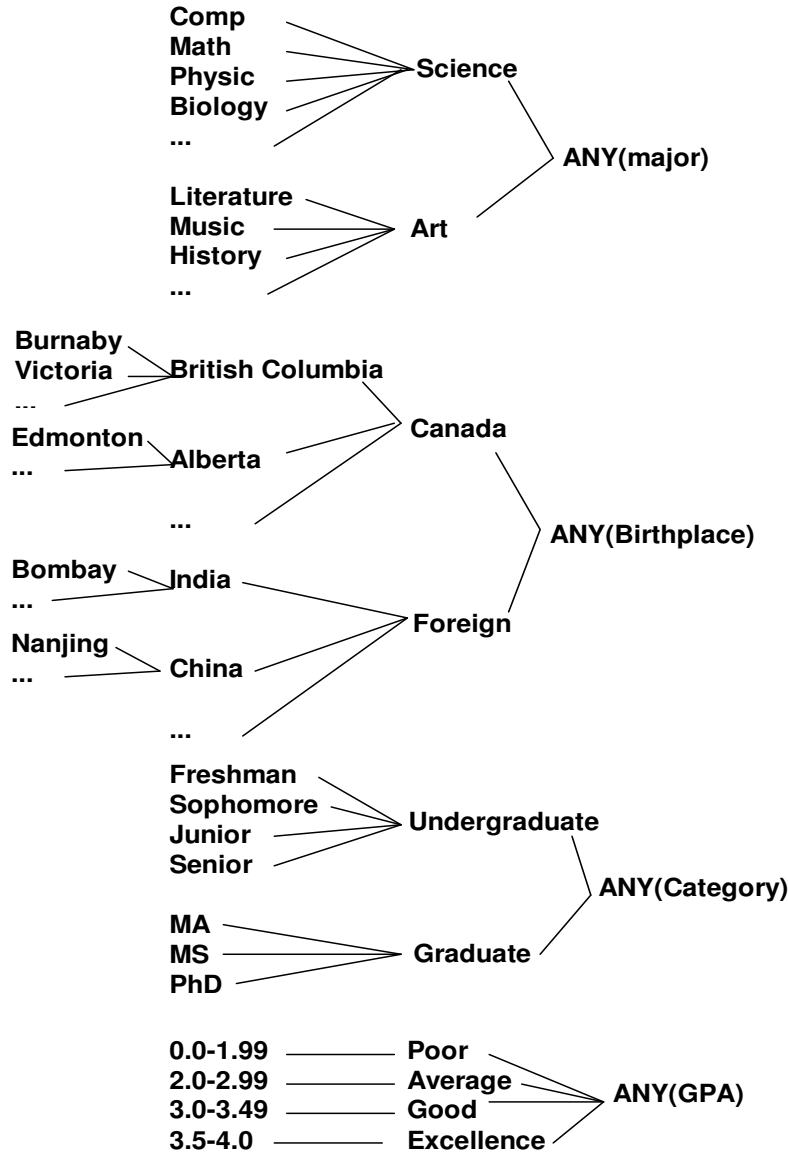

Figure 3. A concept hierarchy table

4. Current attribute oriented induction characteristic rule implementation

In this section will be explained anything regarding with the characteristic rule implementation for current attribute oriented induction where using the current generalization strategy steps (Cai 1989; Han et al 1992) and current characteristic rule algorithm (Cai 1989; Han et al 1992). For doing the generalization there are 8 strategy steps must be done (Han et al. 1992), where step 1 until 7 as for characteristic rule and step 1 until 8 as for classification/discriminant rule. The generalization strategy steps are:
(1) Generalization on the smallest decomposable components
(2) Attribute removal
(3) Concept tree ascension
(4) Vote propagation
(5) Threshold control on each attribute
(6) Threshold control on generalized relations

(7) Rule transformation
(8) Handling overlapping tuples

The characteristic rule algorithm for the implementation will refer to Han et al. (1992) where the characteristic rule algorithm itself as the implementation of generalization strategy steps from number 1 until 6 and will be implemented with java program. The database will be used MySQL and the database architecture will have the same architecture like DBMiner database architecture in figure 1 where table graduate student in table 1 will be saved in table data and concept hierarchy in figure 3 will be saved in table concept hierarchy. Table 2 is a structure database for table student and table 3 is a structure database for table concept hierarchy.

| Field Name | Type |
|---|---|
| Name | Char(20) |
| Category | Char(15) |
| Major | Char(15) |
| Birthplace | Char(20) |
| GPA | Float |

Table 2. Student structure database

| Field Name | Type |
|---|---|
| Field1 | Char(30) |
| Field2 | Char(30) |

Table 3. Concept hierarchy structure database

Threshold number as a control for maximum number of tuples of the target class in the final generalized relation (Han et al, 1992) in many ranges will be input in running program as an experience. Figure 4 shows the result when the program was run by inputting generalization threshold with 1. Figure 5 shows the result when the program was run by inputting generalization threshold with 2, but the final generalization result did not fulfill the generalization strategy step for threshold control on generalized relations where the number of tuples of a generalized relation in the target class is larger than the generalization threshold value, then further generalization should be performed (Han et al.,1992). Further generalization will be done based on selected attributes and merging of identical tuples and the size of generalization relation will be reduced (Han et al,1992). In target class generalization result, except vote attribute the other 3 attributes which have generalization in concept hierarchy will be used as selected attribute for further generalization.

Figure 6 as a result where major attribute as a selected attribute for further generalization. Figure 7 as a result where birthplace attribute as a selected attribute for further generalization, but because the number of tuples still greater than generalization threshold then the number of tuples will be reduced by unioning based on selected other attributes. Figure 8 as the result where the unioning based on major attribute and figure 9 as the result for unioning based on GPA attribute. Figure 10 as a result where GPA attribute as a selected attribute for further generalization, but because the number of tuples still greater than generalization threshold then the number of tuples will be reduced by unioning based on selected other attributes. Figure 11 as the result where the unioning based on major attribute and figure 12 as the result for unioning based on birthplace attribute.

International Journal of Database Management Systems ( IJDMS ) , Vol.2, No.2, May 2010

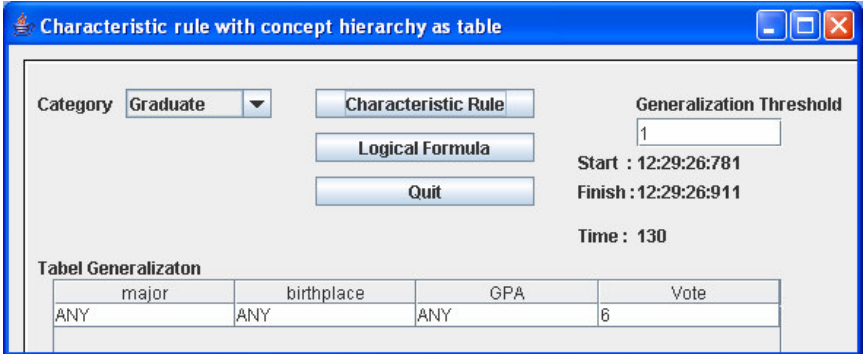

Figure 4. Characteristic rule for current attribute oriented induction program with generalization threshold = 1

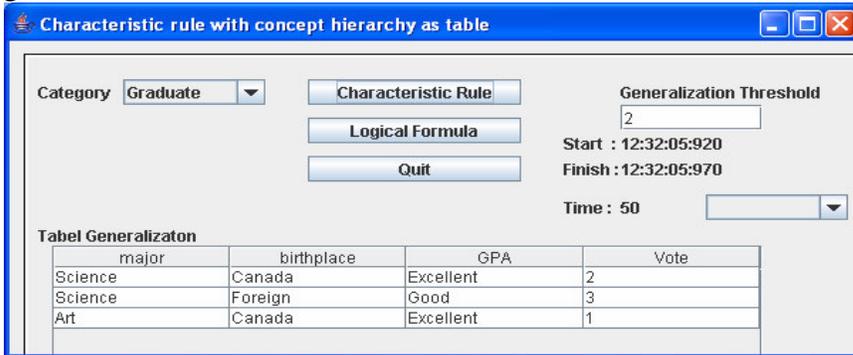

Figure 5. Characteristic rule for current attribute oriented induction program with generalization threshold = 2

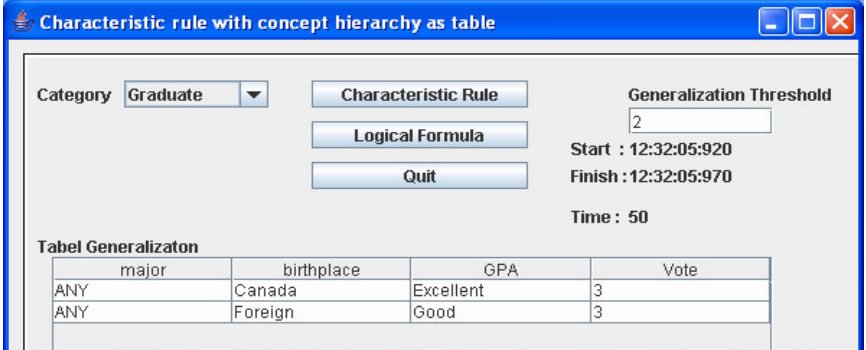

Figure 6. Characteristic rule for current attribute oriented induction program with generalization threshold = 2 and further generalization on major attribute

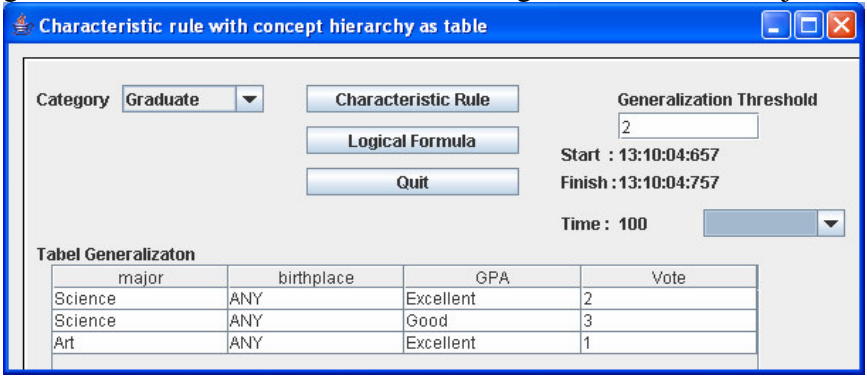

25

Figure 7. Characteristic rule for current attribute oriented induction program with generalization threshold = 2 and further generalization on birthplace attribute

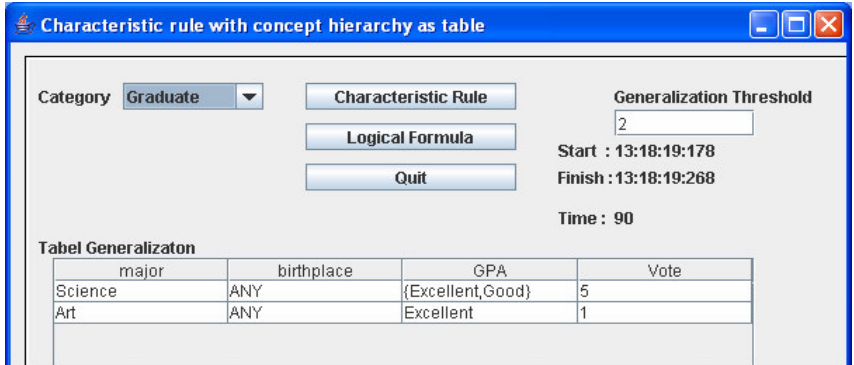

Figure 8. Characteristic rule for current attribute oriented induction program with generalization threshold = 2, further generalization on birthplace attribute and unioning on major attribute

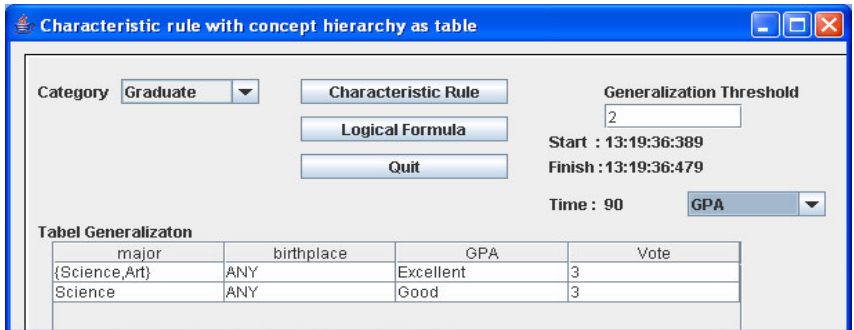

Figure 9. Characteristic rule for current attribute oriented induction program with generalization threshold = 2, further generalization on birthplace attribute and unioning on GPA attribute

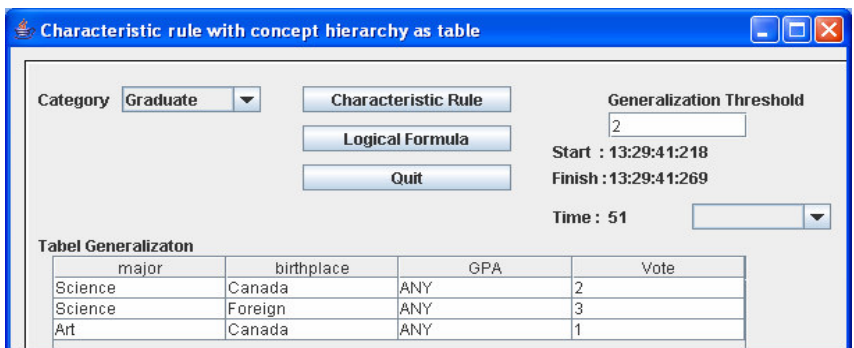

Figure 10. Characteristic rule for current attribute oriented induction program with generalization threshold = 2 and further generalization on GPA attribute

International Journal of Database Management Systems ( IJDMS ) , Vol.2, No.2, May 2010

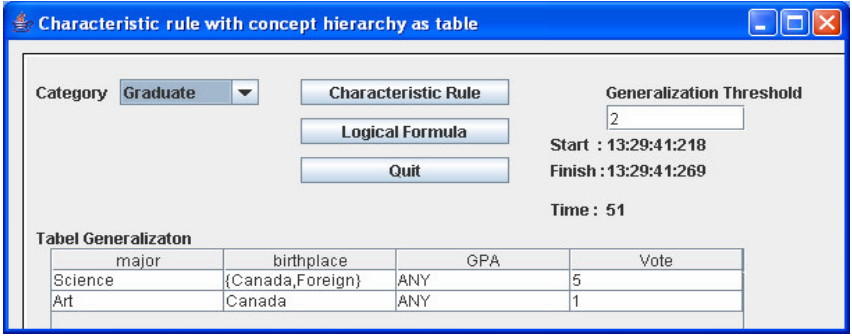

Figure 11. Characteristic rule for current attribute oriented induction program with generalization threshold = 2, further generalization on GPA attribute and unioning on major attribute

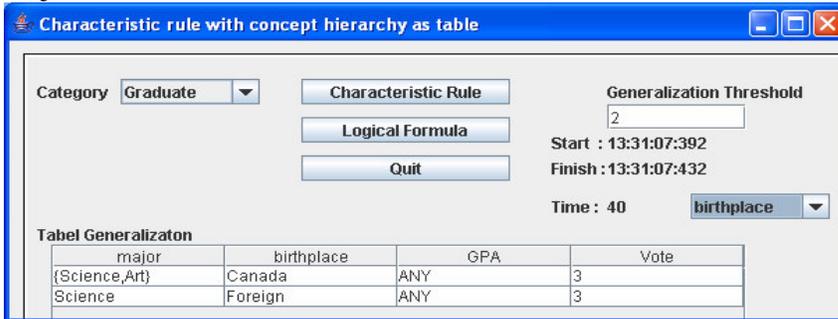

Figure 12. Characteristic rule for current attribute oriented induction program with generalization threshold = 2, further generalization on GPA attribute and unioning on birthplace attribute

Figure 13 shows the result when the program was run by inputting generalization threshold with 3 and Figure 14 shows the result when the program was run by inputting generalization threshold with 4 but because the final generalization result did not fulfill the generalization strategy step for threshold control on generalized relations where the number of tuples of a generalized relation in the target class is larger than the generalization threshold value, then further generalization should be performed (Han et al.,1992). The same as before further generalization will be done based on selected attributes and merging of identical tuples and the size of generalization relation will be reduced (Han et al,1992). In target class generalization result, except vote attribute the other 3 attributes which have generalization in concept hierarchy will be used as selected attribute for further generalization.

Figure 15 as a result where major attribute as a selected attribute for further generalization and Figure 16 as a result where birthplace attribute as a selected attribute for further generalization. Figure 17 as a result where GPA attribute as a selected attribute for further generalization, but because the number of tuples still greater than generalization threshold then the number of tuples will be reduced by unioning based on selected other attributes. Figure 18 as the result where the unioning based on major attribute and figure 19 as the result for unioning based on birthplace attribute.

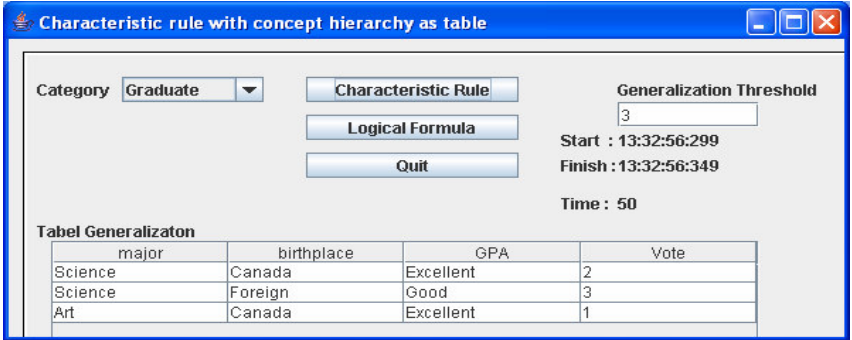

Figure 13. Characteristic rule for current attribute oriented induction program with generalization threshold = 3

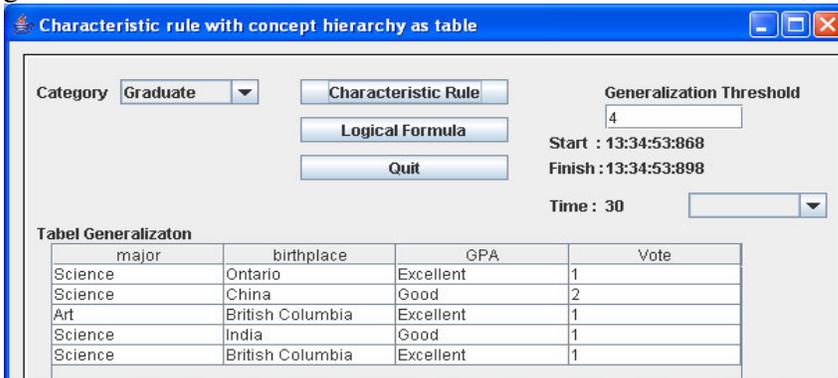

Figure 14. Characteristic rule for current attribute oriented induction program with generalization threshold = 4

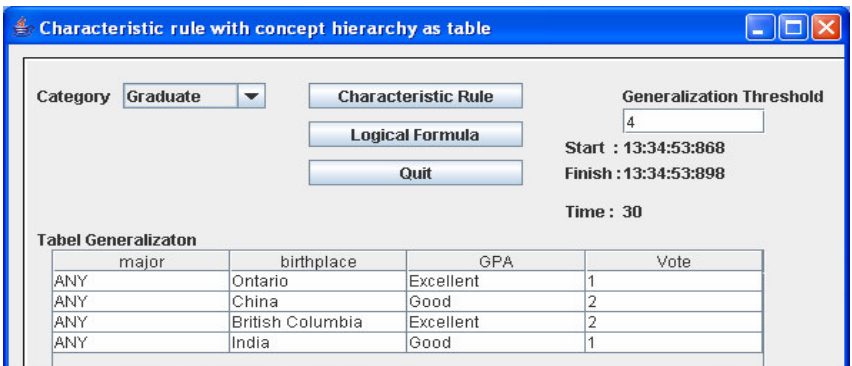

Figure 15. Characteristic rule for current attribute oriented induction program with generalization threshold = 4 and further generalization on major attribute

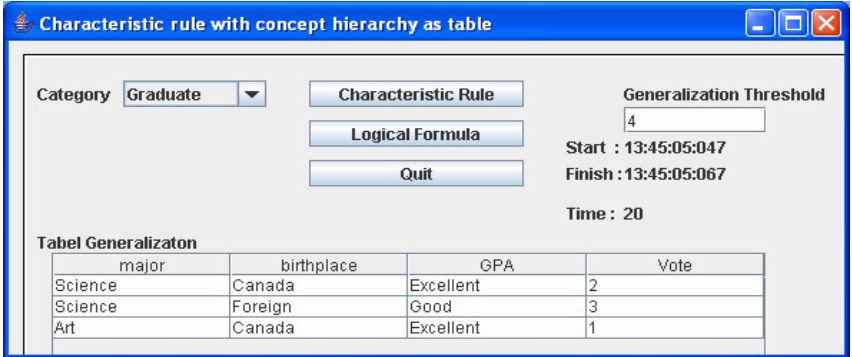

Figure 16. Characteristic rule for current attribute oriented induction program with generalization threshold = 4 and further generalization on birthplace attribute

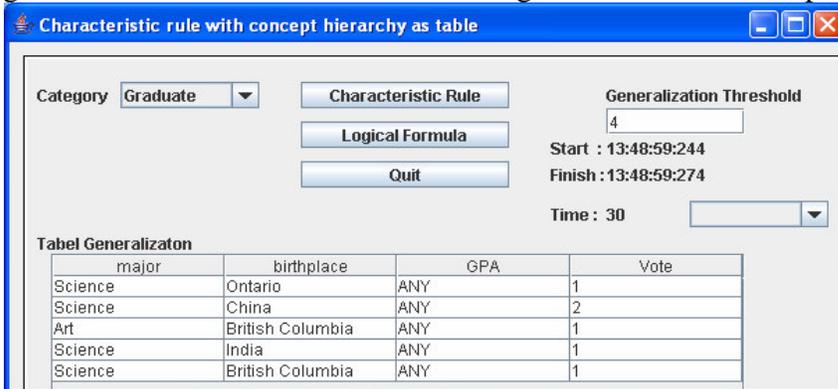

Figure 17. Characteristic rule for current attribute oriented induction program with generalization threshold = 4 and further generalization on GPA attribute

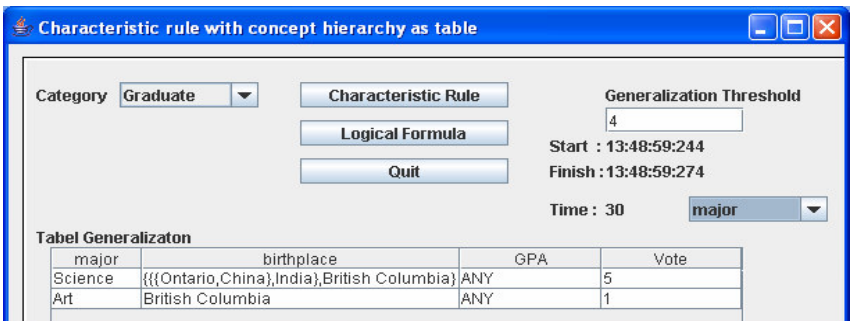

Figure 18. Characteristic rule for current attribute oriented induction program with generalization threshold = 4, further generalization on GPA attribute and unioning on major attribute

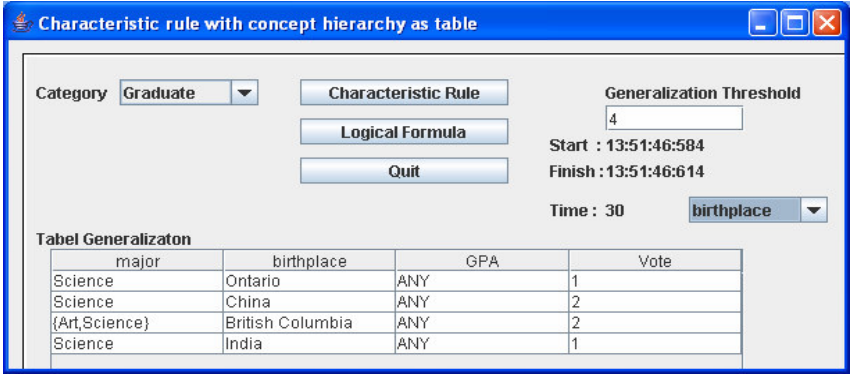

Figure 19. Characteristic rule for current attribute oriented induction program with generalization threshold = 4, further generalization on GPA attribute and unioning on birthplace attribute

Figure 20 shows the result when the program was run by inputting generalization threshold with 5 and Figure 21 shows the result when the program was run by inputting generalization threshold with 6 or more.

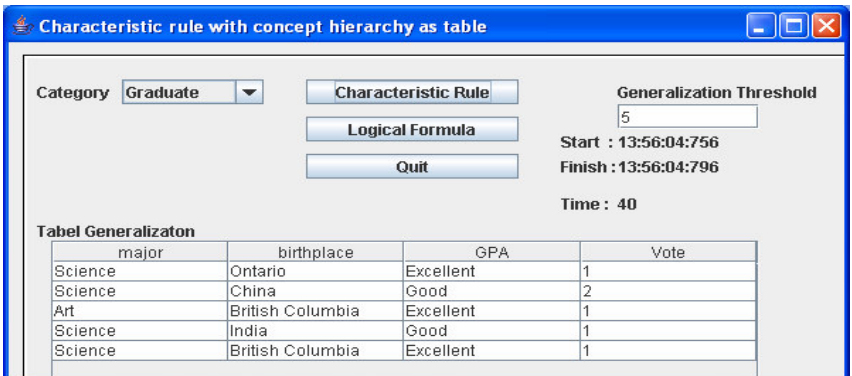

Figure 20. Characteristic rule for current attribute oriented induction program with generalization threshold = 5

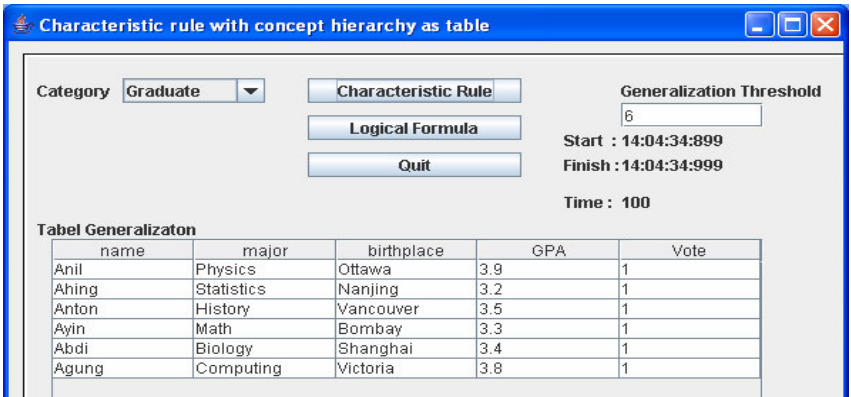

Figure 21. Characteristic rule for current attribute oriented induction program with generalization threshold = 6 or more

5. Star schema attribute induction characteristic rule implementation

International Journal of Database Management Systems ( IJDMS ) , Vol.2, No.2, May 2010

In this section will be explained the uniqueness regarding with the characteristic rule implementation for novel approach star schema attribute induction. Star schema attribute induction has uniqueness like:
(1) Elimination ANY as the most general concept.
(2) Replacement the role concept hierarchy as saving generalization background knowledge with concept tree.
(3) Elimination threshold number as maximum tuples control for generalization result.
(4) Simplification the generalization strategy steps
(5) Elimination attribute oriented induction algorithm

Attribute without higher level concept like attribute name will make consequence the attribute will be removed from process learning and there is other possibility to generalize the value to ANY and then remove the attribute (Han et al. 1992; Han et al. 1993) because of ANY or null description does not provide interesting information on the attribute, over generalization rule and lost some valuable information (Cai, 1989). In star schema attribute induction, ANY as the most general concept will be eliminated and figure 22 shows concept tree which is adopted from figure 3 but without ANY as the most general concept. Program implementation will be implemented with java program, using MySQL database and the database architecture as shown in figure 2.

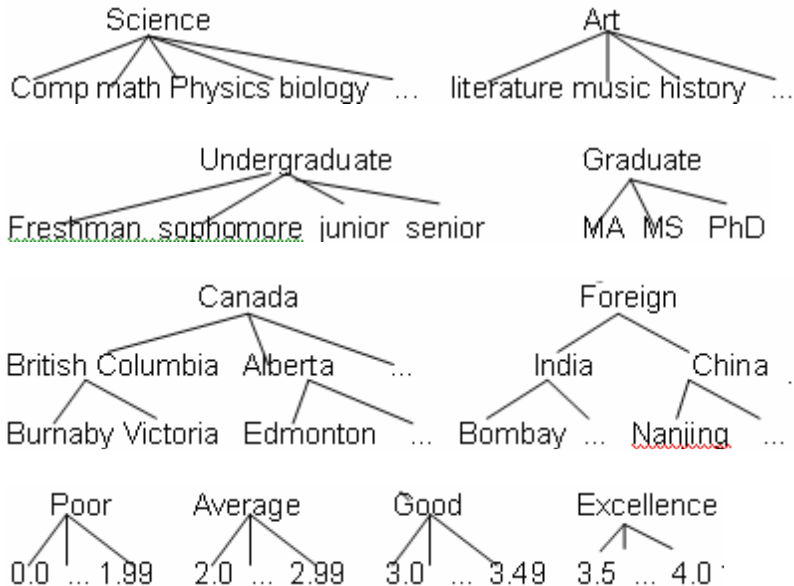

Figure 22. Concept tree without ANY

The same like current attribute oriented induction table graduate student in table 1 will be saved in table data and table 2 is a structure database for table student. In current attribute oriented induction the role of concept hierarchy as saving generalization background knowledge will be replaced concept tree in star schema attribute induction. The amount of concept tree tables will depend on the amount concept trees in concept hierarchy. Structure database for each of concept tree table will depend for each of concept tree data type. Next transformation concept tree in figure 22 into structure table database will be explained.

Concept tree for major in figure 22 will be implemented as table hierarchy_major in table 4 as the transformation is explained in figure 23. The lowest level concept tree for major in figure 22 like comp, math, physics, biology, literature, music and history become the first field name Major and has varchar data type with length 15. The next and the last level concept tree for major in figure 2 like science and art become the next field name StudyProg and have varchar data type with length 15.

| Field Name | Type |
|---|---|
| Major | varchar(15) |
| StudyProg | varchar(15) |

Table 4. table hierarchy_major

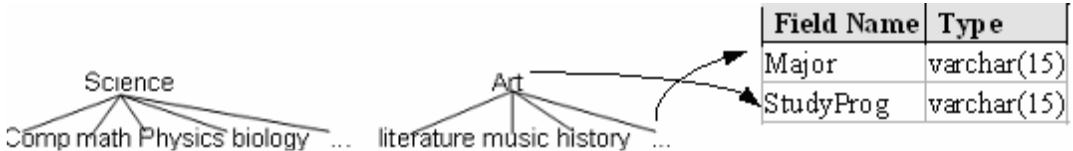

Figure 23. Transformation concept tree for Major into table hierarchy_major

For example there are 10 the lowest level concepts from concept tree major in figure 22 which will create 10 records or tuples in table hierarchy_major based on field Major as the first field in table hierarchy_major. Each of record will fill the next field studyprog based on generalization the lowest level concept tree for major. For example the record where the major field was filled with computing will fill the next field studyprog with science because the concept computing as the lowest level concept tree major in figure 22 has generalization into Science concept. As the result table 5 is the data from concept tree for major in figure 22.

| Major | StudyProg |
|---|---|
| Computing | Science |
| Math | Science |
| Biology | Science |
| Chemistry | Science |
| Statistics | Science |
| Physics | Science |
| Music | Art |
| History | Art |
| Literal Arts | Art |
| Literature | Art |

Table 5. Records for table hierarchy_major

Concept tree for category in figure 22 will be implemented as table hierarchy_cat in table 6 as the transformation is explained in figure 24. The lowest level concept tree for major in figure 22 like Fresman, Sophomore, Junior, Senior, MS, MA and PhD become the first field name Category and has varchar data type with length 15. The next and the last level concept tree for category in figure 22 like Undergraduate and Graduate become the next field name Study and has varchar data type with length 15.

| Field Name | Type |
|---|---|

| Category | varchar(15) |
|---|---|
| Study | varchar(15) |

Table 6. table hierarchy_cat

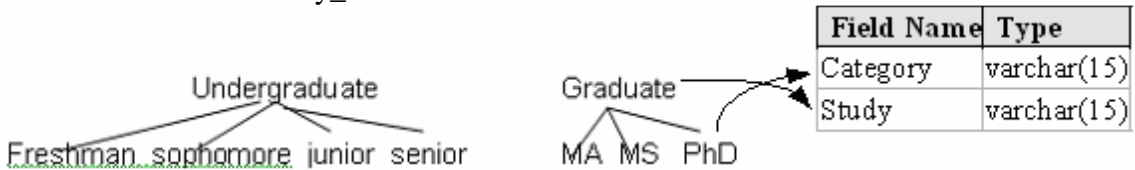

Figure 24. Transformation concept tree for Category into table hierarchy_cat

There are 7 the lowest level concepts from concept tree category in figure 22 which will create 7 records or tuples in table hierarchy_cat based on field Category as the first field in table hierarchy_cat. Each of record will fill the next field Study based on generalization the lowest level concept tree for category. For example the record where the major field was filled with Freshman will fill the next field Study with undergraduate because the concept Freshman as the lowest level concept tree category in figure 22 has generalization into undergraduate concept. As the result table 7 is the data from concept tree for category in figure 22.

| Category | Study |
|---|---|
| Freshman | undergraduate |
| Sophomore | undergraduate |
| Junior | undergraduate |
| Senior | undergraduate |
| MS | graduate |
| MA | graduate |
| PhD | graduate |

Table 7. Records for table hierarchy_cat

Concept tree for birthplace in figure 22 will be implemented as table hierarchy_birth in table 8 as the transformation is explained in figure 25. The lowest level concept tree for major in figure 22 like Burnaby, Victoria, Edmonton, Bombay and Nanjing become the first field name Birthplace and has varchar data type with length 15. The next level concept tree for birthplace in figure 22 like British Columbia, Alberta, India and China become the next field name City and has varchar data type with length 20. Different with previous concept trees this concept tree has 3 leveling, the next and the last level concept tree for birthplace in figure 22 like Canada and Foreign become the next field name Country and has varchar data type with length 10. Thus amount of hierarchy leveling in concept tree will decide the quantity fields which will be created. With the previous concept tree, because there are 2 hierarchy leveling then have been created 2 fields for each concept tree and because concept tree birthplace has 3 hierarchy leveling then automatically will be created 3 fields for the table.

| Field Name | Type |
|---|---|
| Birthplace | varchar(15) |
| City | varchar(20) |

| Country | varchar(10) |
|---|---|

Table 8. table hierarchy_birth

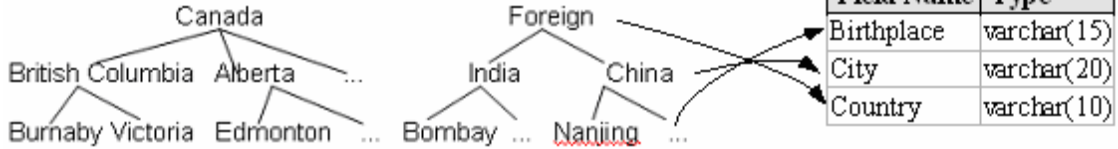

Figure 25. Transformation concept tree for Birthplace into table hierarchy_birth

For example there are 11 the lowest level concepts from concept tree birthplace in figure 22 which will create 11 records or tuples in table hierarchy_birth based on field Birthplace as the first field in table hierarchy_birth. Each of record will fill the next fields based on generalization on concept tree for birthplace. For example the record where the major field was filled with Bombay will fill the next field City with India because the concept Bombay as the lowest level concept tree birthplace in figure 22 has generalization into India concept and last will fill the next and the last field Country with Foreign because the concept India has generalization into Foreign concept. As the result table 9 is the data from concept tree for birthplace in figure 22.

| Birthplace | City | Country |
|---|---|---|
| Bombay | India | Foreign |
| Burnaby | British Columbia | Canada |
| Calgary | Alberta | Canada |
| Edmonton | Alberta | Canada |
| Nanjing | China | Foreign |
| Ottawa | Ontario | Canada |
| Richmond | British Columbia | Canada |
| Shanghai | China | Foreign |
| Toronto | Ontario | Canada |
| Vancouver | British Columbia | Canada |
| Victoria | British Columbia | Canada |

Table 9. Records for table hierarchy_birth

Concept tree for GPA in figure 22 will be implemented as table hierarchy_gpa in table 10 as the transformation is explained in figure 26. Different with the previous concept trees, there are a huge range data for hierarchy leveling, for example the generalization for concept Poor come from range value between 0 and 1.99 and there will be 199 values start from 0.00 until 1.99. For making efficiency then we just record first range values and last range values for each of hierarchy leveling and as the result will add one field. Because concept tree GPA in figure 22 has 2 leveling then should be created 2 fields for the table as we agree before that amount of hierarchy leveling decide the quantity fields will be created, but because for efficiency then the concept trees for numeric values will be treated differently (Han and Fu, 1994; Huang and Lin, 1996; Hu; 2003; Hsu, 2004). Efficiency can be made where only 4 records be created with 3 fields rather than with 400 records with 2 fields for the table.

The first field name GPA_start will filled with first range value and has float(3,2) data type, the next field name GPA_fin will filled with last range value and has float(3,2) data type and the last field name range as the same with other concept trees where the highest level before the most general concept ANY is the last field. The field range has varchar data type with length 15.

| Field Name | Type |
|---|---|
| GPA_start | float(3,2) |
| GPA_fin | float(3,2) |
| range | varchar(15) |

Table 10. table hierarchy_gpa

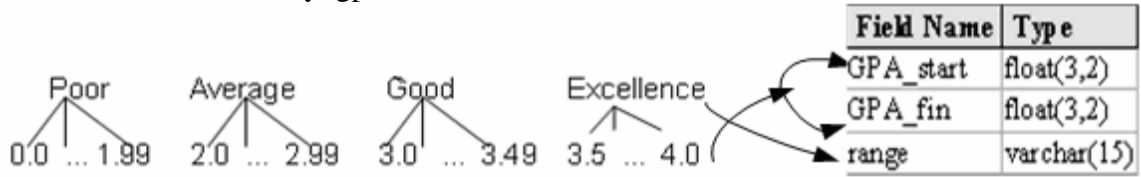

Figure 26. Transformation concept tree for GPA into table hierarchy_gpa

Because concept tree GPA in figure 22 handle for numeric value then the amount of created record will not depend on the amount of concept at the lowest level in concept tree, but the amount of concept at the next generalization as efficiency. At the next generalization after the lowest level concept there are 4 concepts, they are Poor, Average, Good, and Excellence. Different with others, handling for numeric value in concept tree will do the specialization where for example the first data at the level after the first low level is Poor and after that specialize into first range value and put in field GPA_start with value 0 and specialize into last range value and put in field GPA_fin with value 1.99 and as well as others. As the result table 11 is the data from concept tree for GPA in figure 22.

| GPA_start | GPA_fin | range |
|---|---|---|
| 0.0 | 1.99 | Poor |
| 2.0 | 2.99 | Average |
| 3.0 | 3.49 | Good |
| 3.5 | 4.0 | Excellent |

Table 11. Records for table hierarchy_gpa

As conclusion there are some assumptions for converting concept tree into table database :
(1) The lowest level concept tree is the first field and the highest level is the last field.
(2) The amount of hierarchy leveling in concept tree will decide the quantity of fields except for numeric value for efficiency.
(3) The amount of concept at the lowest level in concept tree will become the amount of records or tuples in table except for numeric value for efficiency.
(4) For efficiency, handling numeric value in concept tree, the amount of created table record will not depend on the amount of concepts at the lowest level in concept tree, but the amount of concepts at the next generalization.
(5) The amount of concept tree will decide the amount table of concept tree.

Figure 27 show class diagram the connectivity between table data and concept tree tables as the same like star schema database architecture in figure 2. Refer to Data Warehouse concept, figure 27 will represent as star schema, where student table as fact table and concept tree tables as dimensional tables. As a result multi dimensional concept in Data Warehouse can be applied where data can be roll up and drill down and data can be viewed in multiple dimensions with concept slice, dice and pivot(Chen et al. 1996; Han et al. 1999). Using aggregate count function and Group by operator in sql select statement will represent the roll up process (Gray et al.,1997; Alves and Belo, 2007).

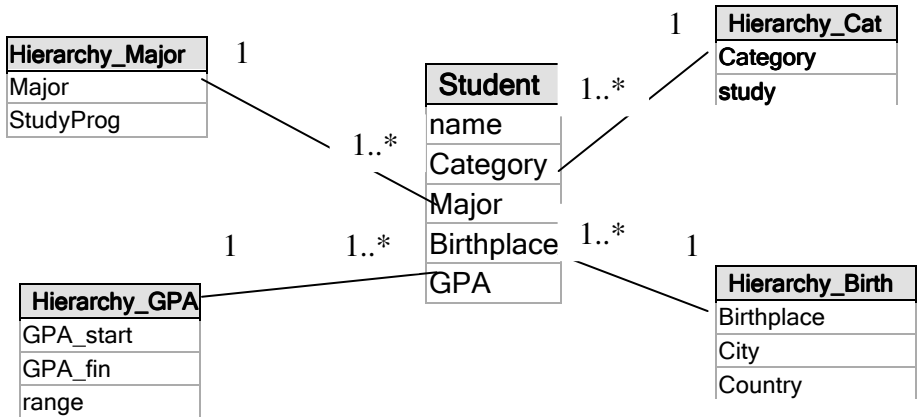

Figure 27. Logical data model star schema attribute induction

Current attribute oriented induction has a weakness where it only provides a snapshot of the generalized knowledge and not a global picture and global picture in current attribute oriented induction can be revealed by trying different thresholds repeatedly (Wu et al 2009). As result by setting different thresholds will obtain different sets of generalized tuples and using different thresholds repeatedly is a time consuming and tedious work (Wu et al 2009). Based on this weaknesses, threshold value as control for maximum number of tuples of target class as learning result is eliminated for star schema attribute induction. Star schema attribute induction does not need control for maximum number of tuples where target class as learning result will be created based on group by operator in sql select statement.

In current attribute oriented induction there are 8 generalization strategy steps as mentioned before and for characteristic rule start from step 1 until 7 they are :
  (1) Generalization on the smallest decomposable components
  (2) Attribute removal
  (3) Concept tree ascension
  (4) Vote propagation
  (5) Threshold control on each attribute
  (6) Threshold control on generalized relations
  (7) Rule transformation
But in star schema attribute oriented induction there are simplification where the generalization strategy steps just only have 3 steps they are :
  (1) Generalization on the smallest decomposable components

(2) Rule transformation

Step 1 in current attribute induction is generalization on the smallest decomposable components where has meaning the generalization should be performed on the smallest decomposable components or attribute of a data relation (Han et al. 1993) or relevant to the learning request (Han et al. 1995b). The process for step 1 in current attribute oriented induction is done by relational query as the implementation of learning task and the same in star schema attribute induction the relational query will prepare the relevant learning request. The next sql statement is the query which is run in java program implementation characteristic rule for star schema attribute induction.

```
select hierarchy_major.studyprog,hierarchy_birth.country,hierarchy_gpa.range,
       count(hierarchy_gpa.range) as uu
from student, hierarchy_cat, hierarchy_major, hierarchy_birth, hierarchy_gpa
where hierarchy_cat.study='"+cmbcategory.getItemAt(cmbcategory.getSelectedIndex())
       and student.major=hierarchy_major.major
       and student.birthplace=hierarchy_birth.birthplace
       and student.category=hierarchy_cat.category
       andstudent.gpa>=hierarchy_gpa.gpa_start
       and student.gpa<=hierarchy_gpa.gpa_fin
group by hierarchy_major.studyprog,hierarchy_birth.country,hierarchy_gpa.range
```

Step 2 in current attribute induction is attribute removal where in star schema attribute induction does not need attribute removal where each of selected attribute will be selected in select statement if there is relation with concept tree table and by other words only attribute in data table which has relation with concept tree table will be selected in select sql statement.

Step 3 in current attribute induction is concept tree ascension where in star schema attribute induction does not need concept tree ascension because selection attributes in select sql statement just only based on attribute which has relation with concept tree table then selection attributes in select sql statement just only for the last field in concept tree table where having role as a highest concept level in concept tree. By choosing the last field in concept tree table for selection attributes in select sql statement have intention to produce the final generalization result consist only small number of tuples and can be simple to be transformed into simple logical formula (Han et al. 1992). By choosing not the last field in concept tree table will produce the huge number tuples of final generalization result. But for multidimensional purpose, choosing not the last field in concept tree table after that will perform drill down process in concept multidimensional.

Step 4 in current attribute induction is vote propagation where vote is accumulated in the generalized relation when merging identical tuples (Han et al. 1992) but in star schema attribute induction vote propagation is accumulated with combination function count( ) and group by operator in select sql statement.

Step 5 in current attribute induction is threshold control on each attribute, because star schema attribute induction does not has threshold control then this step will be eliminated and the same for step 6 in current attribute induction as threshold control on generalized relations. Step 7 in current attribute induction is rule transformation where both of approaches need this step to transform final generalization result become simple logical formula. More than that the 8$^{th}$ step generalization strategy will not need anymore because query operation for preparing data learning request has been extended to produce the final generalization result. As a result both of characteristic rule and classification rule in star schema attribute induction will have the same generalization strategy steps.

Because of there is simplification in the generalization strategy steps and elimination threshold control in star schema attribute induction then algorithm for current attribute induction will be eliminated because query operation for preparing data learning request has been extended to produce the final generalization result in star schema attribute induction. Elimination algorithm is not just only for characteristic rule but also for classification rule in star schema attribute induction because both of them have the same generalization strategy steps and have the same query operation formula in preparing data learning request.

Figure 28 is a result when the star schema attribute induction program is run and the result have the same result in figure 13 for characteristic rule in current attribute oriented induction program with threshold value = 3. For making final generalization result become simple and consist only small number of tuples and can be transformed into simple logical formula then unioning will be done based on each of attribute. Figure 29 is a result for unioning on major attribute and figure 30 is a result for unioning on birthplace or GPA attribute. Unioning on birthplace or GPA attribute can have the same result because there is redundant data like birthplace=Canada and GPA=excellent.

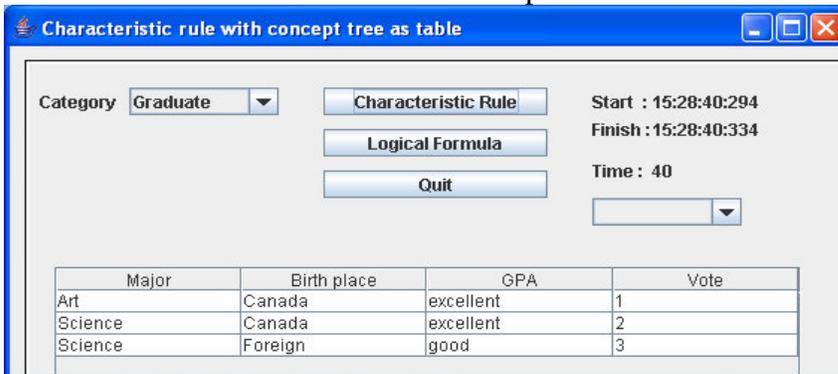

Figure 28. Characteristic rule for star schema attribute induction program

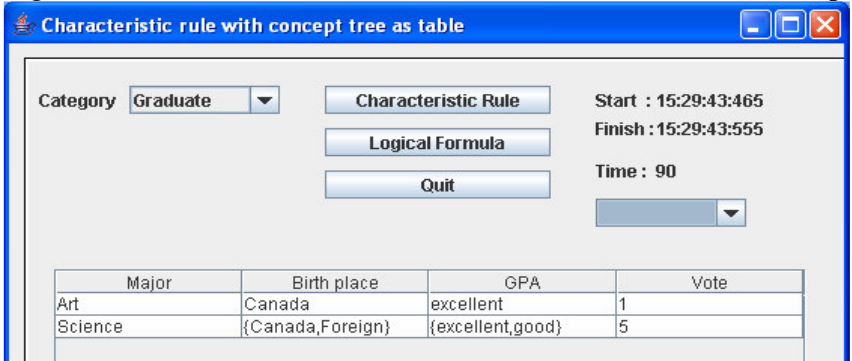

Figure 29. Characteristic rule for star schema attribute induction program with unioning on major attribute

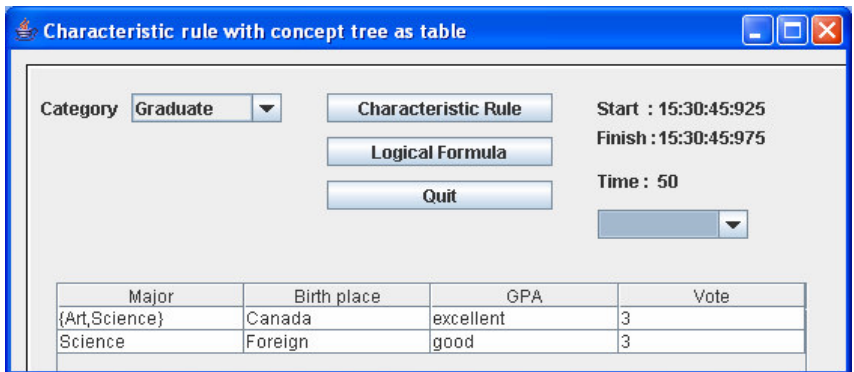

Figure 30. Characteristic rule for star schema attribute induction program with unioning on birthplace or GPA attribute.

6. The differentiation

After get the explanation from both of approaches then we will justify the different as for evident for improvement. Based on the results which were run by each of program application then there are some conclusions:
   (1) Current attribute oriented induction program
      a. Result in generalization threshold = 1 is useless since there are ANY.
      b. Results in generalization threshold = 2, generalization optional and unioning are not optimal since there are ANY in the results.
      c. Results in generalization threshold = 3 are not optimal since can not produce the final generalization result which consist small number of tuples.
      d. Results in generalization threshold = 4 are not optimal since can not produce the final generalization result which consist small number of tuples and since there are ANY in the results.
      e. Results in generalization threshold = 5 are not optimal since can not produce the final generalization result which consist small number of tuples.
      f. Results in generalization threshold = 6 or more are useless since there is no generalization.
   (2) Star schema attribute induction program
      The results as in figure 29 and 30 show that there is no ANY and produce the small number final generalization tuples.

As mentioned before because ANY or null description does not provide interesting information on the attribute, over generalization rule and lost some valuable information (Cai, 1989) and the final generalization result which consist small number of tuples can be simple to be transformed into simple logical formula (Han et al. 1992) then star schema attribute induction have evidence as more powerful than current attribute oriented induction.

The result for performance benchmark, for each of program has been tested based on produce the same result where for current attribute oriented induction with generalization threshold = 3 will produce as in figure 13 and for star schema attribute induction will produce as in figure 28. Both results in figure 13 and figure 28 have the same results and amount of tuples, attributes dan vote value have the same result as well. Timer as benchmarking will start for each of application when Characteristic rule button is pushed and stop when the result is displayed. Both of application programs had been run on the computer with specification Mobile Intel Pentium 4, 2.20 GHz, 644 MHz and 512 MB of RAM and for compatibility each of programs had been run when CPU usage is 1%. Test running program show that current attribute oriented induction program and star schema attribute induction program have the same average performance around 60 milliseconds.

But the problem for star schema attribute oriented induction is where the more concept tree than the more concept tree table must be created and the more using many tables for join operation in query operation will decrease the query performance. In other hands in current attribute oriented induction, increasing of concept hierarchy will not decrease the performance significantly.

7. Conclusion

The differentiation between current attribute oriented induction and star schema attribute induction are:
  (1) The amount of concept tree as simplification hierarchy in current attribute oriented induction can be detected by finding records which have the most general point ANY in the attribute but in star schema attribute induction the amount of concept tree can be detected by the amount of concept tree table.
  (2) The most general concept is the null description described as ANY in current attribute oriented induction is recognized in the attribute on concept hierarchy table but in star schema attribute induction there is no the most general concept and as a result there is no ANY in the attribute on concept tree table.
  (3) The background knowledge in current attribute oriented induction is implemented with one table as concept hierarchy table but in star schema attribute induction is implemented with one or more table as concept tree table.
  (4) Control for maximum number of tuples of the target class in the final generalized relation in current attribute oriented induction is limited by threshold number but in star schema attribute induction is limited by group by operator in sql select statement.
  (5) Using query language in current attribute oriented induction just only happen in the beginning process for collecting the relevant set of data by processing a transformed relational but in star schema attribute induction will be enhanced to generalizes the data by attribute oriented induction.
  (6) There are 7 generalization strategy steps for characteristic rule and 8 generalization strategy steps for classification rule in current attribute oriented induction, but in star schema attribute induction have been simplified for just only 2 generalization strategy steps for both of characteristic rule and classification rule.
  (7) Using algorithm in current attribute oriented induction but in star schema attribute induction the algorithm has been eliminated both of characteristic rule and classification rule. This is because query language for preparing data learning request has been extended to produce the final generalization result.

The disadvantages of star schema attribute induction is the more concept trees the more table concept tree must be created and as implication will increase the amount of tables for join operation in sql statement and automatically will decrease query performance.